\documentclass[a4paper, oneside, twocolumn, notitlepage, 10pt]{extarticle_ecoc}
\usepackage{ecoc}
\usepackage{amsmath,amssymb,amsfonts}
\usepackage{algorithmic}
\usepackage{graphicx}
\usepackage{textcomp}
\usepackage{xcolor}
\usepackage{KITcolors}

\usepackage{pgfplots}
\pgfplotsset{width=10cm,compat=1.9}
\usepackage{caption}
\usepackage{subcaption}

\newcommand{\Blocal}{\boldsymbol{H}_\text{local}}
\newcommand{\Bleft}{\boldsymbol{H}_\text{left}}
\newcommand{\Bright}{\boldsymbol{H}_\text{right}}
\newcommand{\dv}{d_\mathtt{v}}
\newcommand{\dc}{d_\mathtt{c}}
\def\blfootnote{\xdef\@thefnmark{}\@footnotetext}

\begin{document}
\selectlanguage{english}    %

\title{A Spatially Coupled LDPC Coding Scheme  with Scalable Decoders for Space Division Multiplexing}%

\author{
    Haizheng Li and Laurent Schmalen
}

\maketitle                  %

\begin{strip}
 \begin{author_descr}
   
   Communications Engineering Lab, Karlsruhe Institute of Technology (KIT), \textcolor{blue}{\uline{haizheng.li@kit.edu}}
   
 \end{author_descr}
\end{strip}

\setstretch{1.1}
\renewcommand\footnotemark{}
\renewcommand\footnoterule{}

\newcommand\extrafootertext[1]{%
    \bgroup
    \renewcommand\thefootnote{\fnsymbol{footnote}}%
    \renewcommand\thempfootnote{\fnsymbol{mpfootnote}}%
    \footnotetext[0]{#1}%
    \egroup
}

\begin{strip}
  \begin{ecoc_abstract}
    In this paper, we study the application of spatially coupled LDPC codes with sub-block locality for space division multiplexing. We focus on the information exchange between the sub-blocks and compare decoding strategies with respect to the complexity, performance and the information flow. %
  \end{ecoc_abstract}
\end{strip}

\section{Introduction}
Low-density parity-check (LDPC) codes are a large family of error correcting codes with good correction capabilities at moderate decoder complexity~\cite{1057683,leven2014status}. Spatial coupling (SC) further improves their performance and provides the possibility to achieve the channel capacity with  ubiquitous belief propagation (BP) decoding~\cite{6589171,5695130} and has proven to be very useful in optical communications~\cite{schmalen2015spatially}. In \cite{9594186,9174265}, spatially coupled (SC) LDPC codes with sub-block locality and their corresponding decoders were proposed for data storage systems and analyzed on a binary erasure channel (BEC). This new class of codes offer random access to every sub-block in an SC-LDPC code with flexible decoding complexity and performance.\extrafootertext{This work has received funding from the German Federal Ministry of Education and Research (BMBF) under grant agreement 16KIS1420 (STARFALL) and the European Research Council (ERC) under the EU's Horizon 2020 research and innovation programme (grant agreement 101001899).} \extrafootertext{This paper is a preprint of a paper submitted to ECOC 2023 and is subject to Institution of Engineering and Technology Copyright. If accepted, the copy of record will be available at IET Digital Library.}

While this class of codes was designed for random access of small data units with low latency in storage systems, it can also be beneficially used in optical communications with space division multiplexing (SDM). SDM is attractive due to the scaling of the data rate with the number of spatial channels (e.g., cores in a multi-core fiber or modes in a multi-mode fiber). To achieve the best possible performance, we jointly encode the different SDM channels and assign each channel a sub-block of the SC-LDPCL code. The architecture of future SDM receivers is not yet fully clear: On the one hand, the receiver may be fully integrated, processing all SDM channels in a single circuit. On the other hand, a single circuit may not be able to handle the high data rate in SDM and the receiver must be distributed to multiple processing units, each decoding a separate channel. In this paper, we use the SC-LDPC codes with sub-block locality to jointly encode different SDM channels and enable various receiver options: joint decoding in an integrated receiver, fully separate decoding with distributed circuits and semi-joint decoding, where the different receiver circuits can exchange a limited amount of information. We propose decoder variants that improve decoding and reduce the information flow between processing units; these are candidates for future, scalable SDM systems.
\begin{figure*}[t]
    \begin{subfigure}{0.33\textwidth}
        \centering
        \begin{tikzpicture}
    \draw[very thick] (2, 1) rectangle (4, 1.7);
    \draw[very thick] (2, -0.35) rectangle (4, 0.35);
    \draw[very thick] (2, -1.7) rectangle (4, -1);
    \node at (3, 1.35) {\small Decoder 1};
    \node at (3, 0) {\small Decoder 2};
    \node at (3, -1.35) {\small Decoder 3};

    \draw[very thick, ->, -latex] (1.3, 1.35) -- (2, 1.35);
    \draw[very thick, ->, -latex] (1.3, 0) -- (2, 0);
    \draw[very thick, ->, -latex] (1.3, -1.35) -- (2, -1.35);
    \node at (1, 1.35) {$ y_1 $};
    \node at (1, 0) {$ y_2 $};
    \node at (1, -1.35) {$ y_3 $};
							
    \draw[very thick, ->, -latex] (4, 1.35) -- (4.7, 1.35);
    \draw[very thick, ->, -latex] (4, 0) -- (4.7, 0);
    \draw[very thick, ->, -latex] (4, -1.35) -- (4.7, -1.35);
    \node at (5, 1.35) {$ \hat{u}_1 $};
    \node at (5, 0) {$ \hat{u}_2 $};
    \node at (5, -1.35) {$ \hat{u}_3 $};
\end{tikzpicture}
        \caption{Separate decoding}
        \label{fig_separate_decoding}
    \end{subfigure}
    \begin{subfigure}{0.33\textwidth}
        \centering
        \begin{tikzpicture}
    \draw[very thick] (-1.1, -1.7) rectangle (1.1, 1.7);
    \node[align=center,font=\small] at (0, 0) {Joint \\ Decoding};
							
    \draw[very thick, ->, -latex] (-1.8, 1.35) -- (-1.1, 1.35);
    \draw[very thick, ->, -latex] (-1.8, 0) -- (-1.1, 0);
    \draw[very thick, ->, -latex] (-1.8, -1.35) -- (-1.1, -1.35);
    \node at (-2.1, 1.35) {$ y_1 $};
    \node at (-2.1, 0) {$ y_2 $};
    \node at (-2.1, -1.35) {$ y_3 $};
							
    \draw[very thick, ->, -latex] (1.1, 1.35) -- (1.8, 1.35);
    \draw[very thick, ->, -latex] (1.1, 0) -- (1.8, 0);
    \draw[very thick, ->, -latex] (1.1, -1.35) -- (1.8, -1.35);
    \node at (2.1, 1.35) {$ \hat{u}_1 $};
    \node at (2.1, 0) {$ \hat{u}_2 $};
    \node at (2.1, -1.35) {$ \hat{u}_3 $};
\end{tikzpicture}
        \caption{Joint decoding}
        \label{fig_joint_decoding}
    \end{subfigure}
    \begin{subfigure}{0.33\textwidth}
        \centering
        \begin{tikzpicture}
    \draw[very thick] (2, 1) rectangle (4, 1.7);
    \draw[very thick] (2, -0.35) rectangle (4, 0.35);
    \draw[very thick] (2, -1.7) rectangle (4, -1);
    \node at (3, 1.35) {\small Decoder 1};
    \node at (3, 0) {\small Decoder 2};
    \node at (3, -1.35) {\small Decoder 3};

    \draw[very thick, ->, -latex] (1.3, 1.35) -- (2, 1.35);
    \draw[very thick, ->, -latex] (1.3, 0) -- (2, 0);
    \draw[very thick, ->, -latex] (1.3, -1.35) -- (2, -1.35);
    \node at (1, 1.35) {$ y_1 $};
    \node at (1, 0) {$ y_2 $};
    \node at (1, -1.35) {$ y_3 $};
							
    \draw[very thick, ->, -latex] (4, 1.35) -- (4.7, 1.35);
    \draw[very thick, ->, -latex] (4, 0) -- (4.7, 0);
    \draw[very thick, ->, -latex] (4, -1.35) -- (4.7, -1.35);
    \node at (5, 1.35) {$ \hat{u}_1 $};
    \node at (5, 0) {$ \hat{u}_2 $};
    \node at (5, -1.35) {$ \hat{u}_3 $};

    \draw[very thick, <->, dashed, -latex] (2.8, 0.35) -- (2.8, 1);
    \draw[very thick, <->, dashed, -latex] (2.8, -0.35) -- (2.8, -1);
    \draw[very thick, <->, dashed, -latex] (3.9, .95) .. controls (4.5, 0.8) and (4.5, -0.8) .. (3.9, -0.95);
    \node at (2,0.65) [align=right,font=\scriptsize]{Information\\ exchange};
\end{tikzpicture}
        \caption{Semi-joint decoding}
        \label{fig_neighbor_decoding}
    \end{subfigure}
    \vspace*{2ex}
    \caption{Decoding modes in an SDM optical communication system investigated in this paper, 3 SDM channels}
    \label{fig_decoding_modes}
\end{figure*}
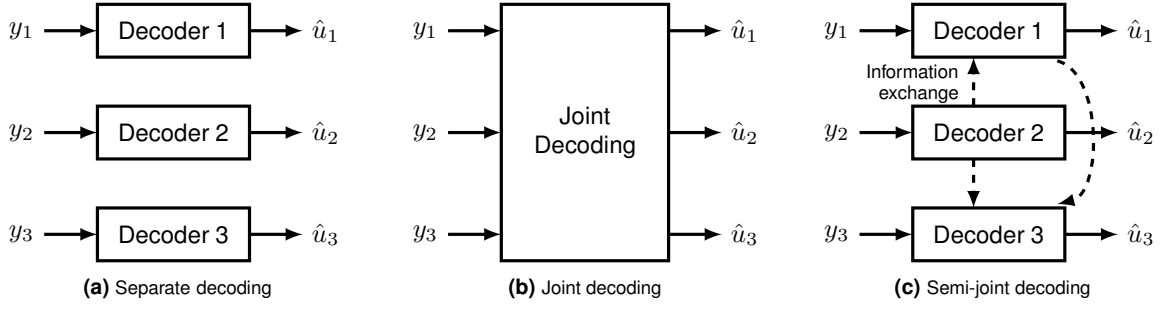\vspace*{-0.7ex}
\section{Background}

A $(\dv, \dc)$ LDPC code is a linear block code given by its parity check matrix $\boldsymbol{H}$, which has $\dv$ 1s in each column and $\dc$ 1s in each row. 
SC-LDPC codes divide the codeword into sub-blocks and the parity check equations (check nodes) 
connect neighboring sub-blocks. In this work, we focus on unit-memory SC-LDPC codes~\cite{schmalen2016design}, where only the directly neighboring sub-blocks contribute to the current sub-block. SC-LDPC codes enable a simple windowed decoding scheme~\cite{schmalen2015spatially}.

The authors in~\cite{9594186,9174265} introduced an additional constraint during the construction of SC-LDPC codes, where a fraction of $t$ check nodes (called \emph{coupled checks}, CC) in a sub-block are allowed to connect to other sub-blocks, and the remaining check nodes (\emph{local checks}, LC)  only connect to code bits in the same sub-block. LCs allow us to decode a single sub-block without information from other sub-blocks; the CCs are the bridge to exchange information between sub-blocks. The parity-check matrix of SC-LDPCL codes is composed of $\Blocal$, which defines the LCs and both $\Bleft$ and $\Bright$, which define the CCs and connections to neighboring sub-blocks. This new SC-LDPCL codes offer three decoding modes: separate local decoding, joint decoding of all sub-blocks, and semi-joint decoding. More details of the decoders can be found in \cite{9594186,9174265}. In the remainder of this paper, we use the $(\dv=4, \dc=20, t=\frac14)$ SC-LDPCL code as an example, due to the good decoding performance of these parameters~\cite{schmalen2015spatially,schmalen2016design}.

We employ the concept of SC-LDPCL codes to jointly encode the spatial channels of SDM. SC-LDPCL codes enable a future-proof system design that allows different, scalable and flexible decoders, depending on the application. We assign each sub-block of the SC-LDPCL code to an SDM channel. In the decoder, we either consider decoding of each sub-block separately  (Fig.~\ref{fig_decoding_modes}(a)) or allow an information exchange between the sub-channels (Fig.~\ref{fig_decoding_modes}(b)), which should be kept as little as possible (Fig.~\ref{fig_decoding_modes}(c)). Our work extends SC-LDPCL codes towards communications. In SDM receivers, we aim at limiting information exchange between decoders, possibly requiring slow communication channels on circuit boards between different decoder circuits. We propose two variants of the semi-joint decoder to cope with different levels of information exchange.
\section{Semi-joint Decoder Variants}
In this section, we first introduce the semi-joint (SJ) decoder and its variants. The SJ decoder is equivalent to the SG decoder from~\cite{9594186,9174265}, extended to communications over general channels. %

\vspace*{0.8ex}
\noindent\emph{SJ Decoder (Conventional SG Decoder)~\cite{8625294}} \\
In SJ decoding, we decode a specific \emph{target} sub-block $T$ with the help of $d$ neighboring sub-blocks, called \emph{helpers}. We assume that the number of involved sub-blocks does not exceed the total number of the sub-blocks and we neglect boundary effects. Furthermore, we assume $d$ to be even, so that the helpers lie symmetrically on both sides of the target. The decoder is described by:\vspace{0.2ex}

\begin{enumerate}
    \item Decode the two furthest sub-blocks $T+\frac d2$ and $\ T-\frac d2$ locally, i.e. carry out BP decoding using the LCs only (effectively using a parity-check matrix $\Blocal$) using the sub-block channel outputs $\boldsymbol{y}_{T- d/2}$ and $\boldsymbol{y}_{T+d/2}$, respectively.
    \item Decode the left helpers: for the $i$-th helper, $T- \frac d2 < i < T$, decode the sub-block with information from its left neighbor, i.e. carry out BP decoding with the parity check matrix $\left[\begin{matrix}\Bright & \Bleft \\ \boldsymbol{0} & \Blocal\end{matrix}\right]$ and $[\hat{\boldsymbol{y}}_{i-1},\ \boldsymbol{y}_i]$, with $\hat{\boldsymbol{y}}_{i-1}$ the updated information of the left neighbor.
    \item Decode the right helpers: for the $i$-th helper, $T<i<T+\frac d2$, decode the sub-block with information from its right neighbor, i.e. do BP decoding with the parity check matrix $\left[\begin{matrix}\Blocal & \boldsymbol{0} \\ \Bright & \Bleft \end{matrix}\right]$ and $[\boldsymbol{y}_i, \hat{\boldsymbol{y}}_{i+1}]$, where $\hat{\boldsymbol{y}}_{i+1}$ is the updated information of the right neighbor.
    \item Decode the target: carry out BP decoding with the parity check matrix $\left[\begin{matrix}\Bright & \Bleft & \boldsymbol{0} \\ \boldsymbol{0} & \Blocal & \boldsymbol{0} \\\boldsymbol{0} & \Bright & \Bleft \end{matrix}\right]$ and $[\hat{\boldsymbol{y}}_{T-1},\ \boldsymbol{y}_T,\ \hat{\boldsymbol{y}}_{T+1}]$.
\end{enumerate}

It should be noted that the decoding of helpers on both sides can be carried out in parallel. In Figs.~\ref{figure_comparison_sg_var1} and \ref{figure_comparison_sg_var2}, the performance of our exemplary code under SJ decoding is given together with the separate and fully joint decoding performance as reference. An increasing number of helpers leads to lower bit error rates (BERs); separate and fully joint decoding are lower and upper bound on the SJ decoding performance. We can see that already for small $d$, SJ decoding closes the significant gap between joint and separate decoding, the latter effectively using a code of higher rate but with complete separate decoders.

\vspace*{0.8ex}
\noindent\emph{SJ Decoder Variant} \\
As stated before, the information flow between the decoder in SJ decoding is an important performance parameter, especially if the decoder are realized in different circuits. In the SJ decoder, we exchange $d$ times  soft information between sub-blocks for decoding a single sub-block. In this variant, we exploit that in an SDM system, the channel information from different sub-channels is available at the same time. The proposed variant is described in what follows:

\begin{enumerate}
    \item For decoding target sub-block $T$, the helpers $T-\frac d2, \dots, T-1$ and $T+1, \dots, T+\frac d2$ transmit their channel information to the target.
    \item Carry out BP decoding using the parity check matrix
    \[
    \boldsymbol{B} = \underbrace{\left[\begin{matrix}
        \Blocal &         &         &         &   \\
        \Bright & \Bleft  &         &         &   \\
                & \Blocal &         &         &   \\
                & \Bright &         &         &   \\
                &         & \ddots  &         &   \\
                &         &         & \Bright & \Bleft \\
                &         &         &         & \Blocal \\
    \end{matrix}\right]}_{d+1 \text{ sub-blocks}}
    \]
    and the channel information
    \[
    [\boldsymbol{y}_{T-\frac d2}, \dots, \boldsymbol{y}_{T-1}, \boldsymbol{y}_T, \boldsymbol{y}_{T+1}, \dots, \boldsymbol{y}_{T+\frac d2}].
    \]
\end{enumerate}
It is easy to show that the complexity of this approach is identical to the SJ decoder and the different decoders only need to exchange information before starting decoding and not during the execution of the decoder. The performance of our example code using the new decoder variant (denoted ``SJVar'') is shown in Fig.~\ref{figure_comparison_sg_var1}. With the same number of helpers $d$, our new variant provides a performance gain of approximately $0.2\mathrm{dB}$, without increasing the information flow between the sub-blocks and the complexity of the decoder.

\begin{figure}[t]
    \begin{tikzpicture}
\begin{semilogyaxis}[
    width=\columnwidth,
    height=0.87\columnwidth,
    xlabel={$ {E_b}/{N_0} $ (dB)},    
    ylabel={BER},
    xmin=2.4, xmax=4.2,
    ymin=1e-5, ymax=1,
    xtick={2.4,2.8,3.2,3.6,4.0},
    ymajorgrids=true,
    xmajorgrids=true,
    legend cell align={left},
    legend style={at={(1, 1)},anchor=north east,font=\scriptsize},
]
\addplot[
    color=black,
    thick,
    mark=triangle,
    mark size = 2pt,
    ]
    coordinates {
        (1.0000e+00, 6.9862e-02)
	(1.1000e+00, 6.7276e-02)
	(1.2000e+00, 6.5031e-02)
	(1.3000e+00, 6.2052e-02)
	(1.4000e+00, 5.9765e-02)
	(1.5000e+00, 5.7273e-02)
	(1.6000e+00, 5.5070e-02)
	(1.7000e+00, 5.2202e-02)
	(1.8000e+00, 4.9320e-02)
	(1.9000e+00, 4.7161e-02)
	(2.0000e+00, 4.4574e-02)
	(2.1000e+00, 4.1737e-02)
	(2.2000e+00, 3.9191e-02)
	(2.3000e+00, 3.6035e-02)
	(2.4000e+00, 3.3411e-02)
	(2.5000e+00, 3.0103e-02)
	(2.6000e+00, 2.7069e-02)
	(2.7000e+00, 2.4229e-02)
	(2.8000e+00, 2.0478e-02)
	(2.9000e+00, 1.6110e-02)
	(3.0000e+00, 1.2031e-02)
	(3.1000e+00, 8.0982e-03)
	(3.2000e+00, 5.2653e-03)
	(3.3000e+00, 3.3317e-03)
	(3.4000e+00, 1.7776e-03)
	(3.5000e+00, 8.7704e-04)
	(3.6000e+00, 3.7363e-04)
	(3.7000e+00, 1.7772e-04)
	(3.8000e+00, 7.6743e-05)
	(3.9000e+00, 3.1091e-05)
    };
\addplot[
    color = KITblue,
    thick,
    mark = x,
    mark size = 2pt,
    ]
    coordinates {
        (1.0000e+00, 7.6194e-02)
	(1.1000e+00, 7.3672e-02)
	(1.2000e+00, 7.1436e-02)
	(1.3000e+00, 6.8222e-02)
	(1.4000e+00, 6.5676e-02)
	(1.5000e+00, 6.2975e-02)
	(1.6000e+00, 5.9938e-02)
	(1.7000e+00, 5.7915e-02)
	(1.8000e+00, 5.4231e-02)
	(1.9000e+00, 5.1477e-02)
	(2.0000e+00, 4.8760e-02)
	(2.1000e+00, 4.5779e-02)
	(2.2000e+00, 4.3592e-02)
	(2.3000e+00, 4.0478e-02)
	(2.4000e+00, 3.6094e-02)
	(2.5000e+00, 3.1505e-02)
	(2.6000e+00, 2.6806e-02)
	(2.7000e+00, 2.1928e-02)
	(2.8000e+00, 1.4898e-02)
	(2.9000e+00, 8.5159e-03)
	(3.0000e+00, 3.8861e-03)
	(3.1000e+00, 1.3650e-03)
	(3.2000e+00, 4.2411e-04)
	(3.3000e+00, 1.2708e-04)
	(3.4000e+00, 4.0581e-05)
    };
\addplot[
    color = KITblue70,
   thick,
    mark = x,
    mark size = 2pt,
    ]
    coordinates {
        (1.0000e+00, 7.5814e-02)
	(1.1000e+00, 7.3532e-02)
	(1.2000e+00, 7.1023e-02)
	(1.3000e+00, 6.8679e-02)
	(1.4000e+00, 6.5573e-02)
	(1.5000e+00, 6.3416e-02)
	(1.6000e+00, 6.0030e-02)
	(1.7000e+00, 5.8115e-02)
	(1.8000e+00, 5.4774e-02)
	(1.9000e+00, 5.2468e-02)
	(2.0000e+00, 4.9378e-02)
	(2.1000e+00, 4.5852e-02)
	(2.2000e+00, 4.3344e-02)
	(2.3000e+00, 3.9542e-02)
	(2.4000e+00, 3.6362e-02)
	(2.5000e+00, 3.2104e-02)
	(2.6000e+00, 2.6958e-02)
	(2.7000e+00, 1.9411e-02)
	(2.8000e+00, 1.1217e-02)
	(2.9000e+00, 4.7691e-03)
	(3.0000e+00, 1.6190e-03)
	(3.1000e+00, 3.3428e-04)
	(3.2000e+00, 5.9536e-05)
	(3.3000e+00, 1.1240e-05)
    };
\addplot[
    color = KITblue50,
    thick,
    mark = x,
    mark size = 2pt,
    ]
    coordinates {
        (1.0000e+00, 7.5977e-02)
	(1.1000e+00, 7.3675e-02)
	(1.2000e+00, 7.1061e-02)
	(1.3000e+00, 6.7954e-02)
	(1.4000e+00, 6.6014e-02)
	(1.5000e+00, 6.2868e-02)
	(1.6000e+00, 6.0234e-02)
	(1.7000e+00, 5.7530e-02)
	(1.8000e+00, 5.4865e-02)
	(1.9000e+00, 5.1838e-02)
	(2.0000e+00, 4.9186e-02)
	(2.1000e+00, 4.6701e-02)
	(2.2000e+00, 4.3018e-02)
	(2.3000e+00, 3.9237e-02)
	(2.4000e+00, 3.5739e-02)
	(2.5000e+00, 3.1615e-02)
	(2.6000e+00, 2.5488e-02)
	(2.7000e+00, 1.7319e-02)
	(2.8000e+00, 8.4575e-03)
	(2.9000e+00, 2.6026e-03)
	(3.0000e+00, 5.2179e-04)
	(3.1000e+00, 5.9924e-05)
	(3.2000e+00, 7.6788e-06)
    };
\addplot[
    color=KITgreen,
    thick,
    mark=square,
    mark size = 2pt,
    ]
    coordinates {
        (1.0000e+00, 7.6256e-02)
	(1.1000e+00, 7.3352e-02)
	(1.2000e+00, 7.0945e-02)
	(1.3000e+00, 6.8168e-02)
	(1.4000e+00, 6.5763e-02)
	(1.5000e+00, 6.3098e-02)
	(1.6000e+00, 6.1118e-02)
	(1.7000e+00, 5.7578e-02)
	(1.8000e+00, 5.4661e-02)
	(1.9000e+00, 5.1983e-02)
	(2.0000e+00, 4.8718e-02)
	(2.1000e+00, 4.6072e-02)
	(2.2000e+00, 4.3479e-02)
	(2.3000e+00, 3.9523e-02)
	(2.4000e+00, 3.5661e-02)
	(2.5000e+00, 3.2378e-02)
	(2.6000e+00, 2.5673e-02)
	(2.7000e+00, 1.8792e-02)
	(2.8000e+00, 1.1036e-02)
	(2.9000e+00, 4.6271e-03)
	(3.0000e+00, 1.4984e-03)
	(3.1000e+00, 3.1579e-04)
	(3.2000e+00, 4.8049e-05)
    };
\addplot[
    color = KITgreen70,
    thick,
    mark = square,
    mark size = 2pt,
    ]
    coordinates {
        (1.0000e+00, 7.6799e-02)
	(1.1000e+00, 7.3450e-02)
	(1.2000e+00, 7.0997e-02)
	(1.3000e+00, 6.8287e-02)
	(1.4000e+00, 6.5930e-02)
	(1.5000e+00, 6.2958e-02)
	(1.6000e+00, 6.0031e-02)
	(1.7000e+00, 5.7693e-02)
	(1.8000e+00, 5.4698e-02)
	(1.9000e+00, 5.2359e-02)
	(2.0000e+00, 4.8998e-02)
	(2.1000e+00, 4.6658e-02)
	(2.2000e+00, 4.2670e-02)
	(2.3000e+00, 3.9248e-02)
	(2.4000e+00, 3.5526e-02)
	(2.5000e+00, 3.1083e-02)
	(2.6000e+00, 2.0336e-02)
	(2.7000e+00, 1.1935e-02)
	(2.8000e+00, 3.9842e-03)
	(2.9000e+00, 9.0912e-04)
	(3.0000e+00, 1.2792e-04)
	(3.1000e+00, 1.1971e-05)
    };
\addplot[
    color = KITgreen50,
    thick,
    mark = square,
    mark size = 2pt,
    ]
    coordinates {
        (1.0000e+00, 7.6443e-02)
	(1.1000e+00, 7.3918e-02)
	(1.2000e+00, 7.0904e-02)
	(1.3000e+00, 6.8296e-02)
	(1.4000e+00, 6.5501e-02)
	(1.5000e+00, 6.2772e-02)
	(1.6000e+00, 6.0240e-02)
	(1.7000e+00, 5.8135e-02)
	(1.8000e+00, 5.5122e-02)
	(1.9000e+00, 5.2581e-02)
	(2.0000e+00, 4.9530e-02)
	(2.1000e+00, 4.6438e-02)
	(2.2000e+00, 4.2858e-02)
	(2.3000e+00, 3.8903e-02)
	(2.4000e+00, 3.5342e-02)
	(2.5000e+00, 2.8832e-02)
	(2.6000e+00, 1.7220e-02)
	(2.7000e+00, 6.7665e-03)
	(2.8000e+00, 1.5441e-03)
	(2.9000e+00, 2.1261e-04)
	(3.0000e+00, 2.5096e-05)
    };
    
\addplot[
    color =KITred,
    thick,
    mark = o,
    mark size = 2pt,
    ]
    coordinates {
        (1.0000e+00, 7.5977e-02)
	(1.1000e+00, 7.3349e-02)
	(1.2000e+00, 7.0835e-02)
	(1.3000e+00, 6.7957e-02)
	(1.4000e+00, 6.5496e-02)
	(1.5000e+00, 6.2752e-02)
	(1.6000e+00, 6.0166e-02)
	(1.7000e+00, 5.7182e-02)
	(1.8000e+00, 5.4519e-02)
	(1.9000e+00, 5.1358e-02)
	(2.0000e+00, 4.8556e-02)
	(2.1000e+00, 4.5596e-02)
	(2.2000e+00, 4.1778e-02)
	(2.3000e+00, 3.8168e-02)
	(2.4000e+00, 3.3148e-02)
	(2.5000e+00, 2.4254e-02)
	(2.6000e+00, 1.4865e-02)
	(2.7000e+00, 3.8504e-03)
	(2.8000e+00, 7.2245e-04)
	(2.9000e+00, 1.1615e-04)
	(3.0000e+00, 1.6211e-05)
    };
\legend{{separate}, {SJ $ d\!=\!2 $}, {SJ $ d\!=\!4 $}, {SJ $ d\!=\!8 $}, {SJVar $d\!=\!2$}, {SJVar $d\!=\!4$}, {SJVar $d\!=\!8$}, {joint}}
\end{semilogyaxis}
\end{tikzpicture} 
    \caption{Performance of $(\dv=4, \dc=20, t=\frac14)$ code under SJ decoding and the proposed variant}
    \label{figure_comparison_sg_var1}
\end{figure}

\vspace*{0.8ex}
\noindent\emph{SJ Decoding with Hard Information Exchange} \\
In both the SJ decoder and the variant, we exchange soft information between the sub-blocks. Soft information is usually quantized using $q$ bits (typically $q = 5,\ldots, 7$). Exchanging hard information significantly reduces the information flow in the system by a factor of $q$. Therefore, we propose a SJ decoder variant, denoted ``SJ-HD'', which allows hard information exchange while degrading the performance only slightly. The idea is that after BP decoding of a helper, the hard decisions of the variable nodes are transmitted along with an estimate $\hat{\delta}_{\mathrm{b}}$ of the BER. Ther BER estimate is obtained from the fraction $\delta_{\mathrm{c}}$  of unfulfilled check equations through
\begin{equation}
    \hat{\delta}_{\mathrm{b}} = \frac12\left(1-\left(1-2\delta_{\mathrm{c}}\right)^{\frac 1\dc}\right)\,.
    \label{equ_BER_estimate}
\end{equation}
The hard decision of the helper $\hat{\boldsymbol{x}}$ and the corresponding BER are then used to calculate soft information (in terms of log-likelihood ratios (LLRs)) for the next stage (next helper or final decoder) via $\boldsymbol{y}=\hat{\boldsymbol{x}}\cdot\ln\left(\frac{1-\hat{\delta}_{\mathrm{b}}}{\hat{\delta}_{\mathrm{b}}}\right)$. 
We compare the the decoding performance with conventional SJ decoding in Fig.~\ref{figure_comparison_sg_var2}. We observe that the reduction of information flow towards hard decision by a factor $q$ is achieved at the cost of only around 0.1\,dB in BER, which offers a reasonable trade-off between information flow and decoding performance.

\begin{figure}[t]
    \begin{tikzpicture}
\begin{semilogyaxis}[
     width=\columnwidth,
    height=0.87\columnwidth,
    xlabel={$ {E_b}/{N_0} $ (dB)},    
    ylabel={BER},
     xmin=2.4, xmax=4.2,
    ymin=1e-5, ymax=1,
    xtick={2.4,2.8,3.2,3.6,4.0},
    ymajorgrids=true,
    xmajorgrids=true,
    legend cell align={left},
    legend style={at={(1, 1)},anchor=north east,font=\scriptsize},
]
\addplot[
    color=black,
   thick,
    mark=triangle,
    mark size = 2pt,
    ]
    coordinates {
        (1.0000e+00, 6.9862e-02)
	(1.1000e+00, 6.7276e-02)
	(1.2000e+00, 6.5031e-02)
	(1.3000e+00, 6.2052e-02)
	(1.4000e+00, 5.9765e-02)
	(1.5000e+00, 5.7273e-02)
	(1.6000e+00, 5.5070e-02)
	(1.7000e+00, 5.2202e-02)
	(1.8000e+00, 4.9320e-02)
	(1.9000e+00, 4.7161e-02)
	(2.0000e+00, 4.4574e-02)
	(2.1000e+00, 4.1737e-02)
	(2.2000e+00, 3.9191e-02)
	(2.3000e+00, 3.6035e-02)
	(2.4000e+00, 3.3411e-02)
	(2.5000e+00, 3.0103e-02)
	(2.6000e+00, 2.7069e-02)
	(2.7000e+00, 2.4229e-02)
	(2.8000e+00, 2.0478e-02)
	(2.9000e+00, 1.6110e-02)
	(3.0000e+00, 1.2031e-02)
	(3.1000e+00, 8.0982e-03)
	(3.2000e+00, 5.2653e-03)
	(3.3000e+00, 3.3317e-03)
	(3.4000e+00, 1.7776e-03)
	(3.5000e+00, 8.7704e-04)
	(3.6000e+00, 3.7363e-04)
	(3.7000e+00, 1.7772e-04)
	(3.8000e+00, 7.6743e-05)
	(3.9000e+00, 3.1091e-05)
    };
\addplot[
    color=KITblue,
   thick,
    mark=x,
    mark size = 2pt,
    ]
    coordinates {
        (1.0000e+00, 7.6194e-02)
	(1.1000e+00, 7.3672e-02)
	(1.2000e+00, 7.1436e-02)
	(1.3000e+00, 6.8222e-02)
	(1.4000e+00, 6.5676e-02)
	(1.5000e+00, 6.2975e-02)
	(1.6000e+00, 5.9938e-02)
	(1.7000e+00, 5.7915e-02)
	(1.8000e+00, 5.4231e-02)
	(1.9000e+00, 5.1477e-02)
	(2.0000e+00, 4.8760e-02)
	(2.1000e+00, 4.5779e-02)
	(2.2000e+00, 4.3592e-02)
	(2.3000e+00, 4.0478e-02)
	(2.4000e+00, 3.6094e-02)
	(2.5000e+00, 3.1505e-02)
	(2.6000e+00, 2.6806e-02)
	(2.7000e+00, 2.1928e-02)
	(2.8000e+00, 1.4898e-02)
	(2.9000e+00, 8.5159e-03)
	(3.0000e+00, 3.8861e-03)
	(3.1000e+00, 1.3650e-03)
	(3.2000e+00, 4.2411e-04)
	(3.3000e+00, 1.2708e-04)
	(3.4000e+00, 4.0581e-05)
    };
\addplot[
    color =KITblue70,
   thick,
    mark = x,
    mark size = 2pt,
    ]
    coordinates {
        (1.0000e+00, 7.5814e-02)
	(1.1000e+00, 7.3532e-02)
	(1.2000e+00, 7.1023e-02)
	(1.3000e+00, 6.8679e-02)
	(1.4000e+00, 6.5573e-02)
	(1.5000e+00, 6.3416e-02)
	(1.6000e+00, 6.0030e-02)
	(1.7000e+00, 5.8115e-02)
	(1.8000e+00, 5.4774e-02)
	(1.9000e+00, 5.2468e-02)
	(2.0000e+00, 4.9378e-02)
	(2.1000e+00, 4.5852e-02)
	(2.2000e+00, 4.3344e-02)
	(2.3000e+00, 3.9542e-02)
	(2.4000e+00, 3.6362e-02)
	(2.5000e+00, 3.2104e-02)
	(2.6000e+00, 2.6958e-02)
	(2.7000e+00, 1.9411e-02)
	(2.8000e+00, 1.1217e-02)
	(2.9000e+00, 4.7691e-03)
	(3.0000e+00, 1.6190e-03)
	(3.1000e+00, 3.3428e-04)
	(3.2000e+00, 5.9536e-05)
	(3.3000e+00, 1.1240e-05)
    };
\addplot[
    color = KITblue,
   thick,
    mark = x,
    mark size = 2pt,
    ]
    coordinates {
        (1.0000e+00, 7.5977e-02)
	(1.1000e+00, 7.3675e-02)
	(1.2000e+00, 7.1061e-02)
	(1.3000e+00, 6.7954e-02)
	(1.4000e+00, 6.6014e-02)
	(1.5000e+00, 6.2868e-02)
	(1.6000e+00, 6.0234e-02)
	(1.7000e+00, 5.7530e-02)
	(1.8000e+00, 5.4865e-02)
	(1.9000e+00, 5.1838e-02)
	(2.0000e+00, 4.9186e-02)
	(2.1000e+00, 4.6701e-02)
	(2.2000e+00, 4.3018e-02)
	(2.3000e+00, 3.9237e-02)
	(2.4000e+00, 3.5739e-02)
	(2.5000e+00, 3.1615e-02)
	(2.6000e+00, 2.5488e-02)
	(2.7000e+00, 1.7319e-02)
	(2.8000e+00, 8.4575e-03)
	(2.9000e+00, 2.6026e-03)
	(3.0000e+00, 5.2179e-04)
	(3.1000e+00, 5.9924e-05)
	(3.2000e+00, 7.6788e-06)
    };
\addplot[
    color=KITorange,
   thick,
    mark=diamond,
    mark size = 2pt,
    ]
    coordinates {
        (1.0000e+00, 7.6760e-02)
	(1.1000e+00, 7.4109e-02)
	(1.2000e+00, 7.1649e-02)
	(1.3000e+00, 6.9396e-02)
	(1.4000e+00, 6.7466e-02)
	(1.5000e+00, 6.4276e-02)
	(1.6000e+00, 6.2066e-02)
	(1.7000e+00, 5.9321e-02)
	(1.8000e+00, 5.6544e-02)
	(1.9000e+00, 5.3416e-02)
	(2.0000e+00, 5.1898e-02)
	(2.1000e+00, 4.8612e-02)
	(2.2000e+00, 4.6115e-02)
	(2.3000e+00, 4.3283e-02)
	(2.4000e+00, 4.0043e-02)
	(2.5000e+00, 3.7128e-02)
	(2.6000e+00, 3.3617e-02)
	(2.7000e+00, 3.0253e-02)
	(2.8000e+00, 2.5674e-02)
	(2.9000e+00, 1.8865e-02)
	(3.0000e+00, 1.2895e-02)
	(3.1000e+00, 7.5084e-03)
	(3.2000e+00, 3.6207e-03)
	(3.3000e+00, 1.4199e-03)
	(3.4000e+00, 4.0319e-04)
	(3.5000e+00, 9.2358e-05)
	(3.6000e+00, 2.0246e-05)
    };
\addplot[
    color = KITorange70,
   thick,
    mark = diamond,
    mark size = 2pt,
    ]
    coordinates {
        (1.0000e+00, 7.6453e-02)
	(1.1000e+00, 7.4895e-02)
	(1.2000e+00, 7.1997e-02)
	(1.3000e+00, 6.8988e-02)
	(1.4000e+00, 6.6798e-02)
	(1.5000e+00, 6.4319e-02)
	(1.6000e+00, 6.1891e-02)
	(1.7000e+00, 5.9296e-02)
	(1.8000e+00, 5.6312e-02)
	(1.9000e+00, 5.4199e-02)
	(2.0000e+00, 5.1740e-02)
	(2.1000e+00, 4.9005e-02)
	(2.2000e+00, 4.6177e-02)
	(2.3000e+00, 4.3083e-02)
	(2.4000e+00, 4.0333e-02)
	(2.5000e+00, 3.7572e-02)
	(2.6000e+00, 3.3875e-02)
	(2.7000e+00, 3.1128e-02)
	(2.8000e+00, 2.4245e-02)
	(2.9000e+00, 1.7215e-02)
	(3.0000e+00, 9.0440e-03)
	(3.1000e+00, 3.9182e-03)
	(3.2000e+00, 9.9590e-04)
	(3.3000e+00, 1.8116e-04)
	(3.4000e+00, 2.0443e-05)
    };
\addplot[
    color = KITorange50,
   thick,
    mark = diamond,
    mark size = 2pt,
    ]
    coordinates {
        (1.0000e+00, 7.6742e-02)
	(1.1000e+00, 7.3913e-02)
	(1.2000e+00, 7.1661e-02)
	(1.3000e+00, 6.9886e-02)
	(1.4000e+00, 6.6876e-02)
	(1.5000e+00, 6.3898e-02)
	(1.6000e+00, 6.1194e-02)
	(1.7000e+00, 5.9181e-02)
	(1.8000e+00, 5.6577e-02)
	(1.9000e+00, 5.3735e-02)
	(2.0000e+00, 5.1571e-02)
	(2.1000e+00, 4.8577e-02)
	(2.2000e+00, 4.6418e-02)
	(2.3000e+00, 4.3017e-02)
	(2.4000e+00, 4.0937e-02)
	(2.5000e+00, 3.7747e-02)
	(2.6000e+00, 3.3709e-02)
	(2.7000e+00, 3.0036e-02)
	(2.8000e+00, 2.3046e-02)
	(2.9000e+00, 1.3057e-02)
	(3.0000e+00, 5.4269e-03)
	(3.1000e+00, 1.3226e-03)
	(3.2000e+00, 1.3522e-04)
	(3.3000e+00, 1.0194e-05)
    };
    \addplot[
    color = black!70,
   thick,
    mark = square,
    mark size = 2pt,
    ]
    coordinates {
        (1.0000e+00, 7.5977e-02)
	(1.1000e+00, 7.3349e-02)
	(1.2000e+00, 7.0835e-02)
	(1.3000e+00, 6.7957e-02)
	(1.4000e+00, 6.5496e-02)
	(1.5000e+00, 6.2752e-02)
	(1.6000e+00, 6.0166e-02)
	(1.7000e+00, 5.7182e-02)
	(1.8000e+00, 5.4519e-02)
	(1.9000e+00, 5.1358e-02)
	(2.0000e+00, 4.8556e-02)
	(2.1000e+00, 4.5596e-02)
	(2.2000e+00, 4.1778e-02)
	(2.3000e+00, 3.8168e-02)
	(2.4000e+00, 3.3148e-02)
	(2.5000e+00, 2.4254e-02)
	(2.6000e+00, 1.4865e-02)
	(2.7000e+00, 3.8504e-03)
	(2.8000e+00, 7.2245e-04)
	(2.9000e+00, 1.1615e-04)
	(3.0000e+00, 1.6211e-05)
    };
\legend{{separate}, {SJ $d\!=\!2$}, {SJ $d\!=\!4$}, {SJ $d\!=\!8$}, {SJ-HD $d\!=\!2$}, {SJ-HD $d\!=\!4$}, {SJ-HD $d\!=\!8$},{joint}}
\end{semilogyaxis}
\end{tikzpicture} 
    \caption{Performance of $(\dv=4, \dc=20, t=\frac14)$ code SJ and HD-SJ decoding}
    \label{figure_comparison_sg_var2}
\end{figure}

Unfortunately, both variants cannot be trivially combined as the SJ decoder uses soft information to get an estimate of the local symbols, while the variant uses soft information from the beginning.
\section{Conclusions}
In this paper, we have discussed the application of SC-LDPCL ensembles for scalable optical communication systems with SDM. We have adapted the SC-LDPCL to the new application scenario and proposed the semi-joint decoder for decoders that are distributed on multiple processors together with two variants: The first variant improves the BER performance without increasing the information exchange, and the second reduces the information exchange significantly with only small performance penalty.

\clearpage
\newpage
\printbibliography

\vspace{-4mm}

\end{document}